\documentclass[a4paper,superscriptaddress]{revtex4}
\usepackage{epsfig}
\usepackage{hyperref}
\begin{document}
\title{Multiband responses in high-T$_c$ cuprate superconductors}
% \author{D.K.~Sunko\thanks{email: dks\@phy.hr}~~and S.~Bari\v si\``c\\
% Department of Physics,\\Faculty of Science,\\ % University of Zagreb,\\
% Bijeni\v cka cesta 32,\\ % HR-10000 Zagreb, Croatia.}
\author{G.~Nik\v{s}i\'{c}}
\affiliation{Department of Physics, Faculty of Science, University of
Zagreb,\\ Bijeni\v cka cesta 32, HR-10000 Zagreb, Croatia.}
\author{I.~Kup\v{c}i\'{c}}
\affiliation{Department of Physics, Faculty of Science, University of
Zagreb,\\ Bijeni\v cka cesta 32, HR-10000 Zagreb, Croatia.}
\author{O.~S.~Bari\v{s}i\'{c}}
\affiliation{Institute of Physics,\\ Bijeni\v cka cesta 54,
HR-10000 Zagreb, Croatia.}
\author{D.~K.~Sunko}\email{dks@phy.hr}
\affiliation{Department of Physics, Faculty of Science, University of
Zagreb,\\ Bijeni\v cka cesta 32, HR-10000 Zagreb, Croatia.}
\author{S.~Bari\v{s}i\'{c}}
\affiliation{Department of Physics, Faculty of Science, University of
Zagreb,\\ Bijeni\v cka cesta 32, HR-10000 Zagreb, Croatia.}
%\date{}
\begin{abstract}

We report on the interplay of localized and extended degrees of freedom in the
metallic state of high-temperature superconductors in a multiband setting.
Various ways in which the bare magnetic response may become incommensurate are
measured against both phenomenological and theoretical requirements. In
particular, the pseudogap temperature is typically much higher than the
incommensurability temperature. When microscopic strong-coupling effects with
real-time dynamics between copper and oxygen sites are included, they tend to
restore commensurability. Quantum transport equations for low-dimensional
multiband electronic systems are used to explain the linear doping
dependence of the dc conductivity and the doping and temperature dependence of
the Hall number in the underdoped LSCO compounds. Coulomb effects of dopands
are inferred from the doping evolution of the Hartree-Fock model parameters.

\vskip 1cm\noindent
\textbf{Key words:} Emery model, magnetic fluctuations, electron-doped
superconductors.

\noindent
\textbf{Running title:} Multiband responses in high-T$_c$ cuprates

\end{abstract}

\maketitle

\section{Introduction}\label{intro}

The mechanism of high-temperature superconductivity (SC) in cuprate
perovskites remains the pre-eminent open problem of solid-state physics today,
more than 25 years after their discovery. A common paradigm of these materials
emphasizes the strong on-site Coulomb repulsion $U$ (``large $U$'') on the
copper sites, based on the ubiquitous insulating antiferromagnetic (AF) phase
of the parent compounds, eventually replaced by the SC phase of the metallic
doped materials. It begs the question, whether the interesting SC part of the
phase diagram is better described starting from a metal, or from an insulator.
A large body of evidence indicates that the metallized materials themselves
suffer a crossover between the ``deeply'' underdoped (conducting without SC)
and ``ordinary'' underdoped (suboptimally SC) regimes, most obviously in the
temperature dependence of the normal-state conductivity, which changes from an
insulating-like upturn, but with finite residual resistivity, to a downturn
interrupted by SC, around 6\% doping e.g.\ in LSCO~\cite{Bollinger11}. It is
also visible in ARPES~\cite{Lee07} as the smoothing out of a fairly
discontinuous Fermi-surface (FS) gap (U-shaped gap) towards a d-wave form
(V-shaped gap). In the normal state, one can see the ARPES intensity around
the nodal point, the so-called FS arc, stretch out towards the zone boundary
as optimal doping is approached.

The unifying theme of this article is that the arc metal behaves essentially
as a Fermi liquid, with two important qualifications. First, the spectral
weight of the conducting states is less than unity, and depends on doping: the
carriers spend part of their time in localized states, because of the large
$U$. Second, the metal cannot be properly described in an effective-mass
approximation, because (a) the pseudogap (PG) scale appears in the low-energy
responses, and (b) the Fermi velocity varies significantly along the arc. Such
a simple understanding of the low-temperature metallic state is contingent,
however, on a multi-band approach, in which the chemical composition of the
copper-oxygen conducting plane plays an important physical role.

\section{Magnetic responses}\label{smoke}

\subsection{The high-temperature approach}

A nearly-universal feature of the underdoped cuprates is an incommensurate
magnetic response, near the AF wave-vector, observed in neutron
scattering~\cite{Reznik04,Dunsiger08,Enoki12}. The incommensurability is
collinear for the underdoped (SC) materials and diagonal for the
deeply-underdoped (non-SC) compositions. When it is collinear, it also changes
to diagonal at large energy-transfer (frequency).

The usual approach~\cite{Qimiao93,Markiewicz10} to modelling such a response
in a metal is to look for an incommensurability in the high-temperature bare
particle-hole ($ph$) susceptibility $\chi^0_{\xi\xi'}$ in the appropriate
excitation channels $\xi$. The general structure of such a susceptibility
is~\cite{OSBarisic12}
\begin{equation}
\chi_{\xi\xi'}=\chi^0_{\xi\xi'}+
\frac{\chi^0_{\xi\alpha}
U_{\mathrm{eff}}\chi^0_{\alpha\xi'}
}
{
1-U_{\mathrm{eff}}\chi^0_{\alpha\alpha}
},
\label{chi}
\end{equation}
where $\alpha$ is the internal degree of freedom presumed to give the
incommensurate response, probed by creating and destroying particles in the
state $\xi$, and with an effective self-interaction (square vertex)
$U_{\mathrm{eff}}$, discussed below. For the copper-oxide planes, $\alpha$ are
the metallic band states projected onto the copper $d$-orbitals. The
high-temperature response can itself be already symmetry-broken~\cite{Kao00},
or bare with renormalized band parameters (HF)~\cite{Markiewicz10}.

\begin{figure}
\includegraphics[height=5cm]{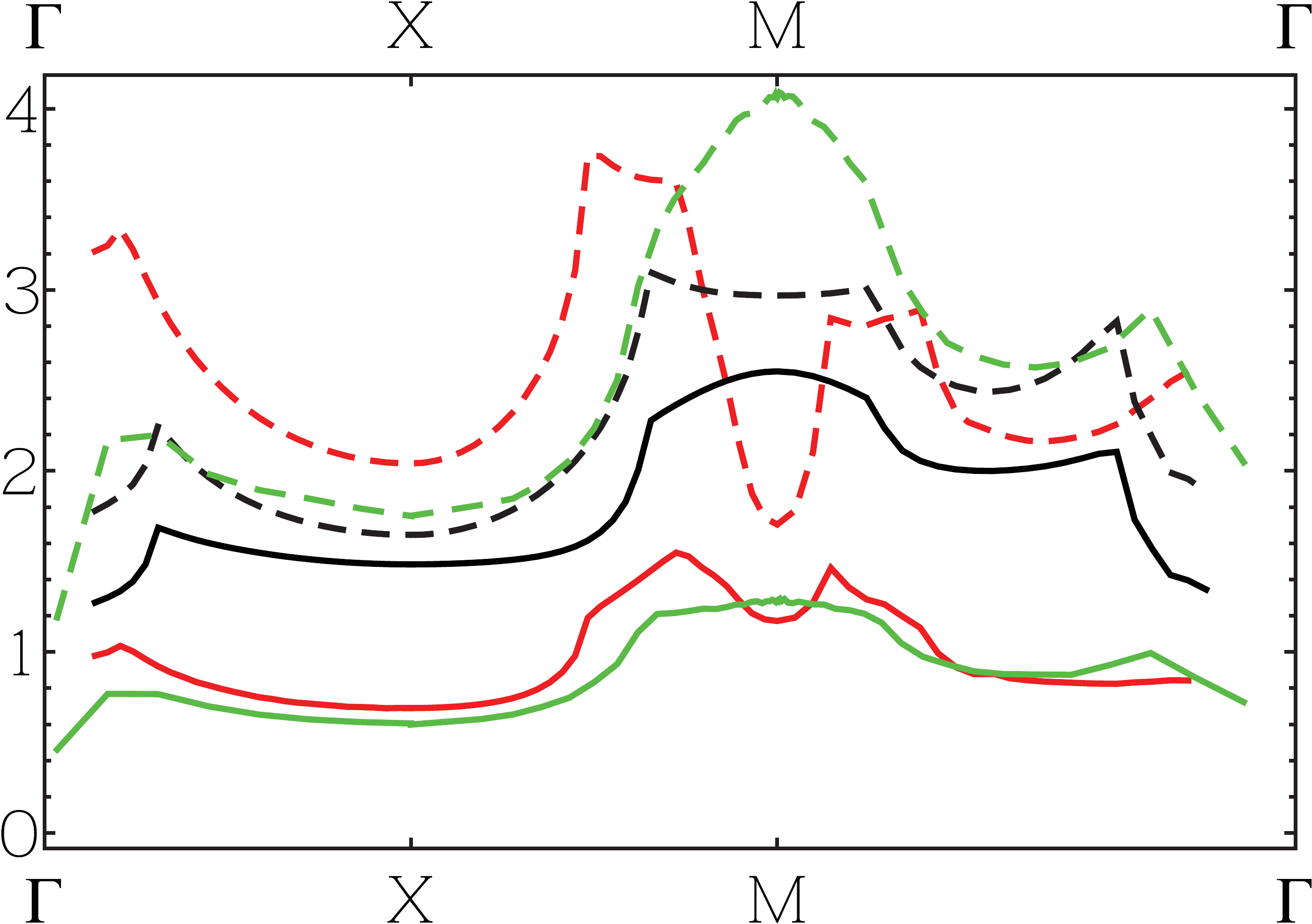}
\includegraphics[height=5cm]{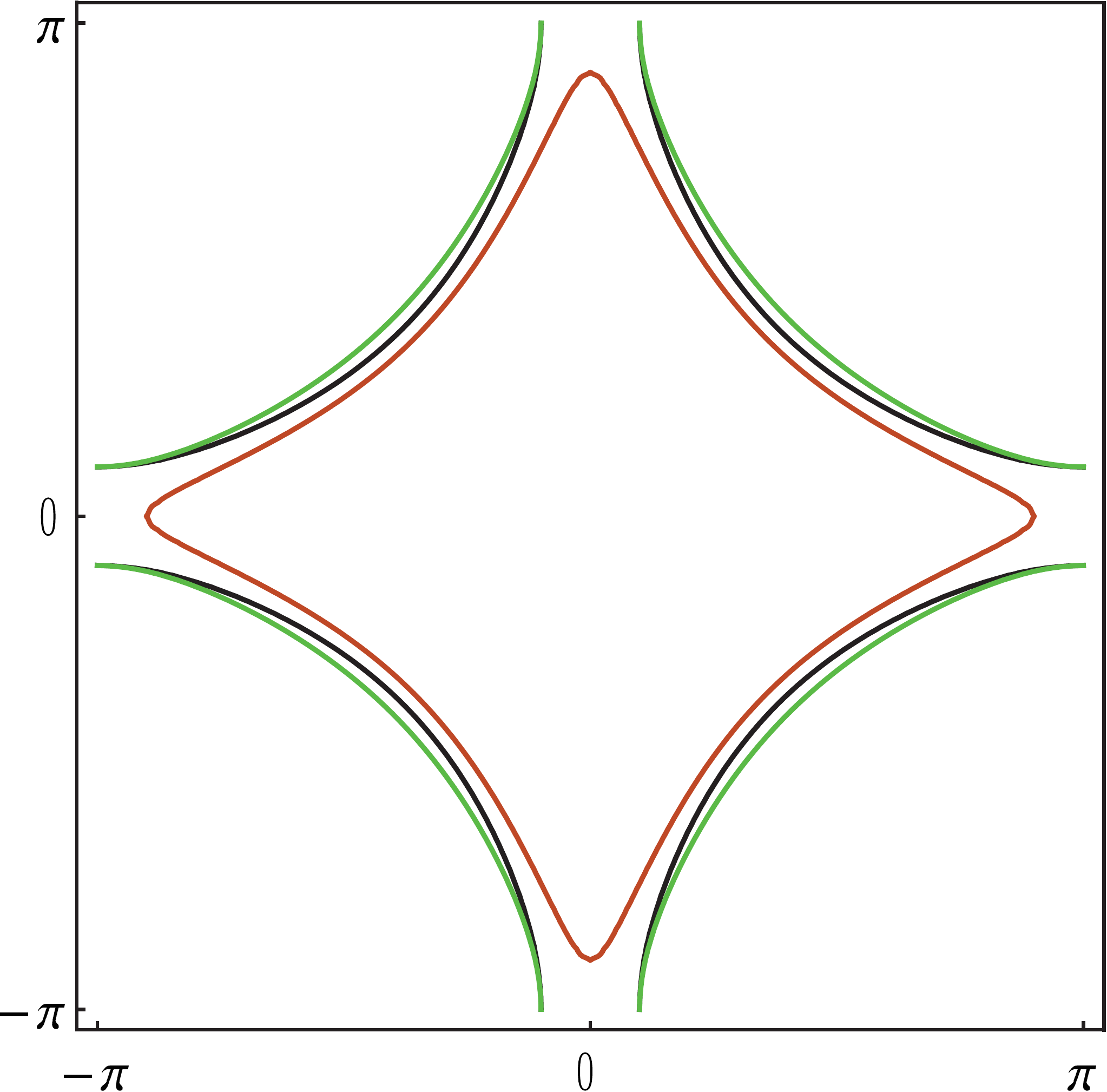}
\caption{Left: free susceptibilities in the three-band model (broken lines),
and with the band-fermion legs renormalized to first order (full lines), as
described in the text. The corresponding Fermi surfaces are shown at right.
The two similar open Fermi surfaces differ in the ratio $|t_{pp}/t_{pd}|$,
which is larger for the one with an incommensurate bare response.}
\label{gama1}
\end{figure}
Our principal result is that the HF response $\chi^0_{\alpha\alpha}$ changes
qualitatively when the strong on-site repulsion is taken into account
dynamically, i.e.\ without any mean-field approximation. Without the on-site
repulsion, regimes can be found for both closed and open Fermi surfaces, where
the responses $\chi^0_{\alpha\alpha}$ are incommensurate.  They are shown by
broken lines in Fig.~\ref{gama1}. Full lines show the corresponding responses
$\chi^1_{\alpha\alpha}$, which include a first-order renormalization of the
metallic band fermions, taking into account the on-site repulsion. Only the
response of the closed (around $\Gamma$) FS remains incommensurate, while all
the open-FS responses are now commensurate.

The large $U$ is infinite in this approach~\cite{OSBarisic12}. The low-energy
scales are generated by a waiting effect, where the metallic hole must wait
for the copper site to be free before visiting it, which creates a scattering
vertex between the hybridized holes in the band. In the low-energy limit, this
gives rise to a constant $U_{\mathrm{eff}}\approx
t_{pd}^4/(\varepsilon_d-\mu)^3\equiv U_{d\mu}$, to be compared with the usual
charge-transfer $J$, where the empty oxygen orbital energy $\varepsilon_p$
appears instead of the band chemical potential $\mu$, leading to similar
numerical scales, $U_{d\mu}\sim J$, despite the profoundly different physics.
Mean-field approaches miss the waiting effect, because they are by
construction only able to renormalize the band parameters, allowing for
qualitatively the same high-temperature incommensurability as in the
non-interacting system, shown by broken lines in the figure. By contrast, the
waiting effect redistributes the spectral strength of the particle-hole
excitations in the Brillouin zone, leading to a qualitatively different,
commensurate susceptibility.

It is also possible to get incommensurability in $\chi_{\xi\xi}$ directly from
the first term, $\chi^0_{\xi\xi}$, by choosing the $\xi$ channel to probe the
oxygen sites, however the response is typically much smaller than
$\chi^1_{\alpha\alpha}$, so it would be quite unnatural to try fine-tuning the
parameters so that the non-interacting term should dominate the interacting
one. Finally, incommensurate responses may appear close enough to half-filling
on both doping sides when $|t_{pp}/t_{pd}|\ll 1$, because the system reverts
to particle-hole symmetry~\cite{Schulz90}. That is not a realistic physical
regime for the cuprates, in which particle-hole symmetry is strongly broken.

The common vein of the standard approaches outlined above is that one gets the
incommensurate response already in the high-temperature phase, which then
triggers an incommensurate low-temperature response through the denominator in
Eq.~(\ref{chi}), at the same wave-vector. Measurements show, however, that the
response is incommensurate at temperatures still above the SC $T_c$, but
considerably below the pseudogap $T^*$, which is the natural candidate for the
transition temperature in this context. Around $T^*$ it is still commensurate,
as our strong-coupling calculation suggests it should be. Therefore we take
the transition from high-temperature to be the commensurate one, and
investigate the so-obtained low-temperature phase in the following.

\subsection{The low-temperature approach}

\begin{figure}
\begin{tabular}{cc}
a)\raisebox{-20mm}{\includegraphics[width=35mm]{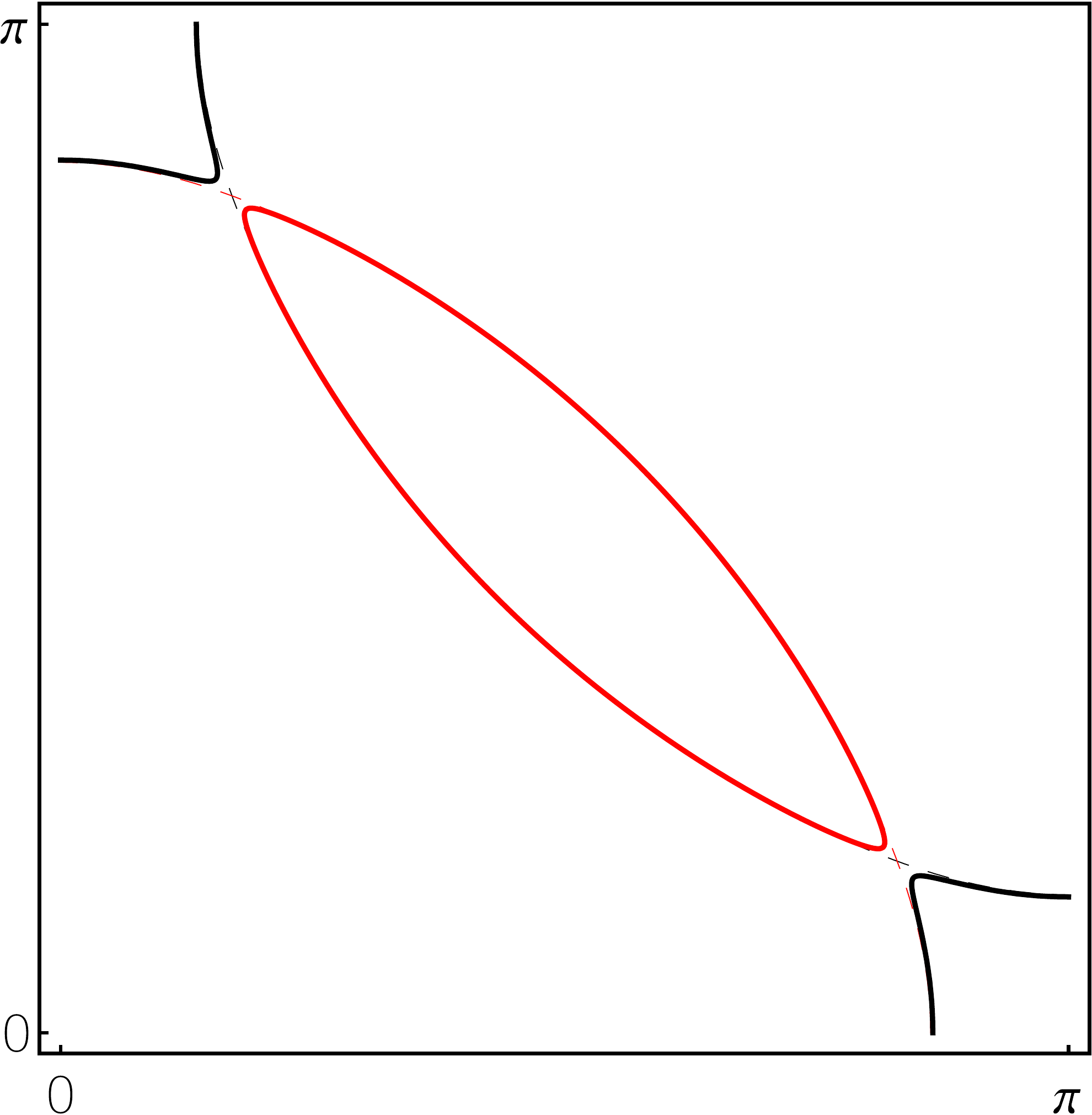}} &
\raisebox{17mm}{b)\includegraphics[width=40mm,angle=-90]{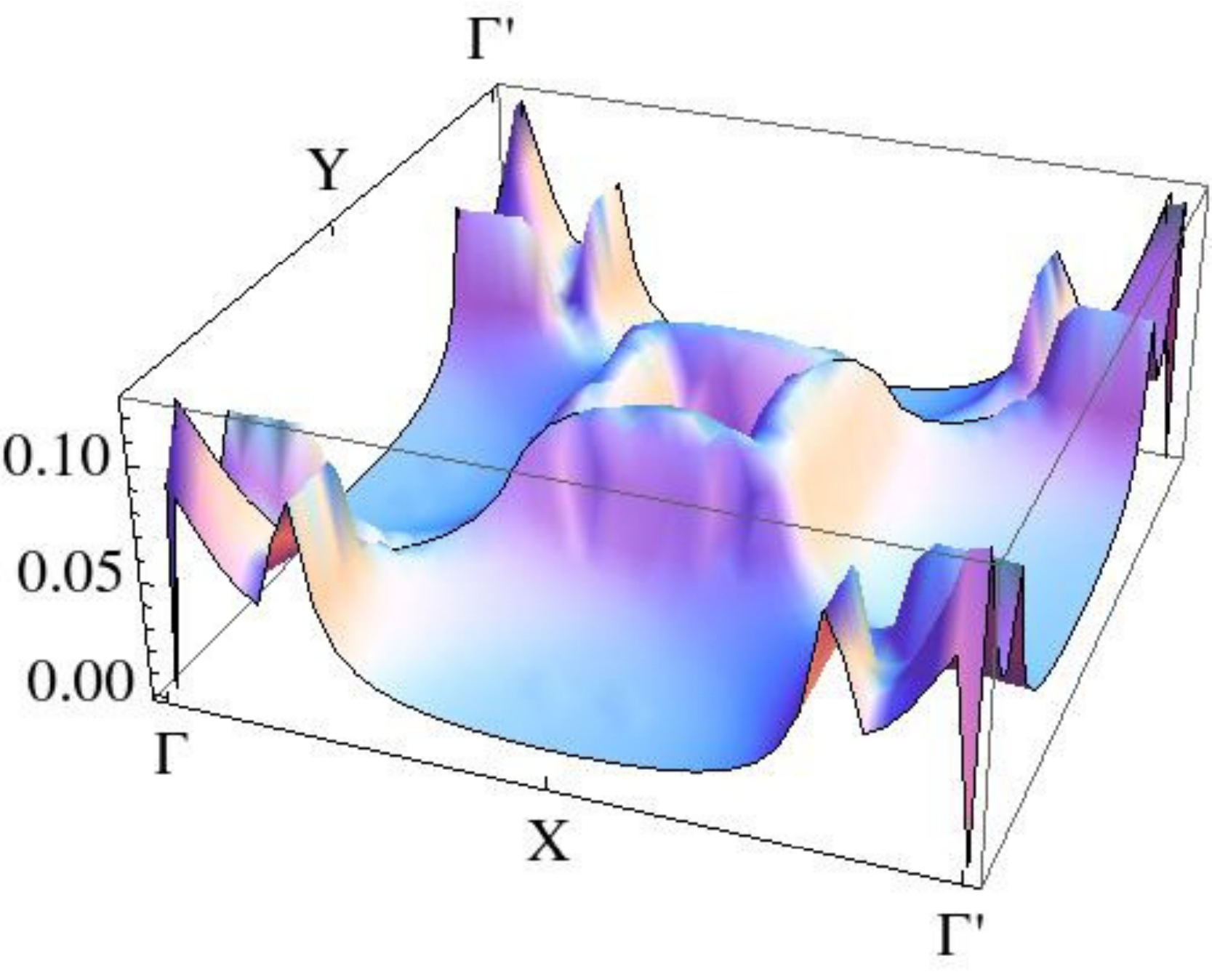}}
\\
\raisebox{17mm}{c)\includegraphics[width=40mm,angle=-90]{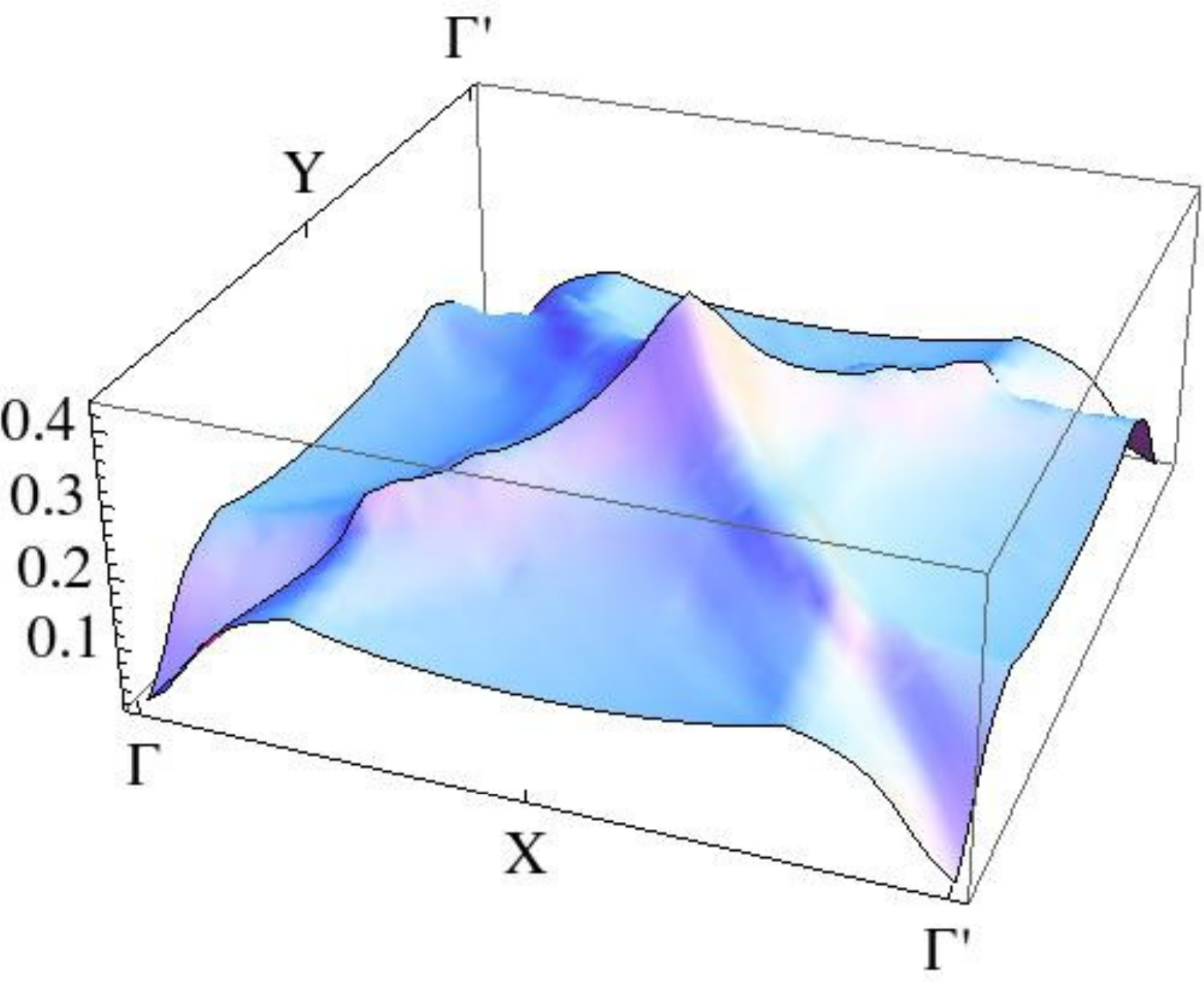}} &
\raisebox{17mm}{d)\includegraphics[width=40mm,angle=-90]{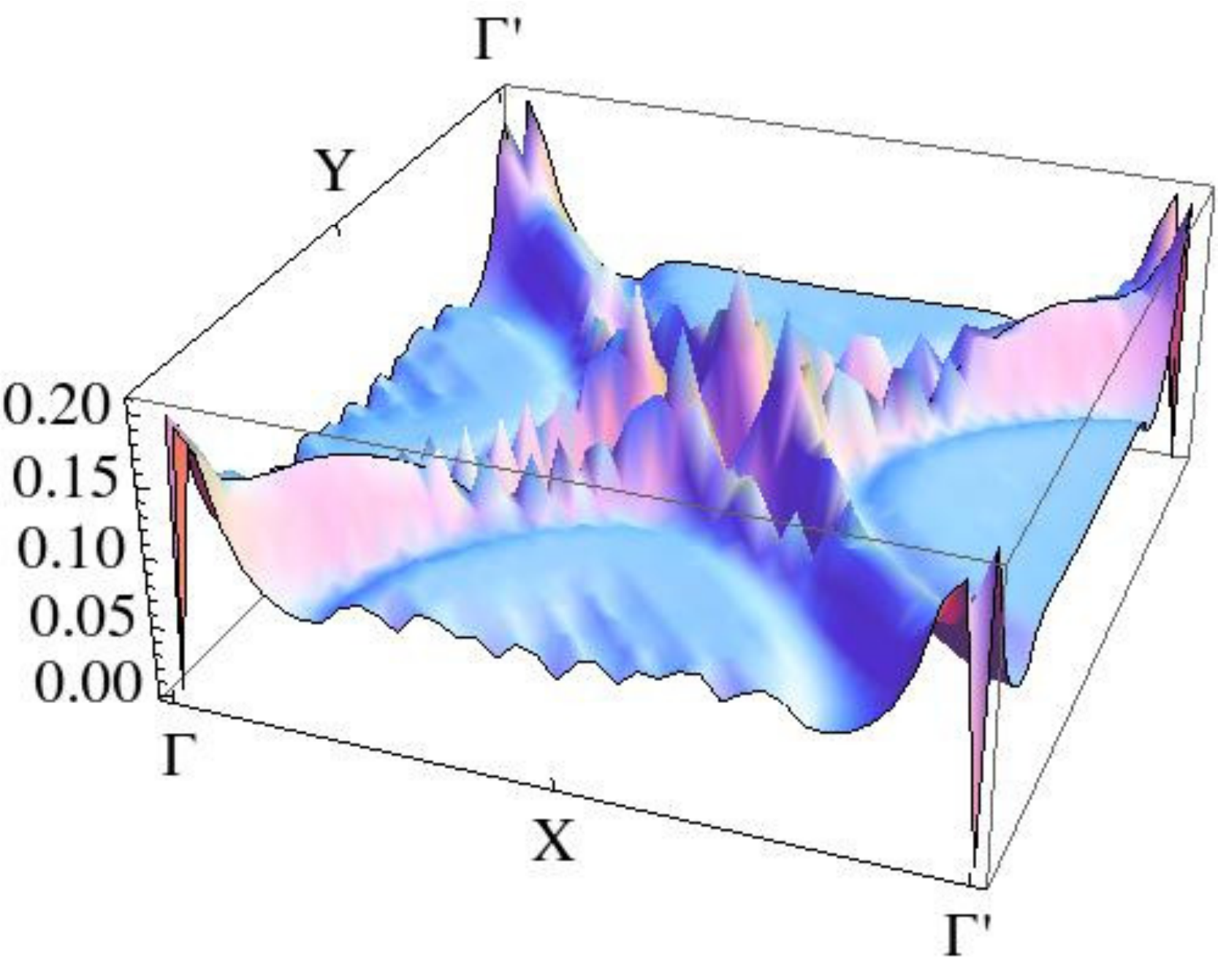}}
\end{tabular}
\caption{a) FS reconstructed from two subbands, showing a nodal and antinodal
part. b) Intraband antinodal response. c) Interband nodal-antinodal response.
d) Intraband nodal response.}
\label{recFS}
\end{figure}
While our high-temperature calculation was fully microscopic in its one
critical feature --- the restoration of commensurability by the dynamical
waiting effect --- the low-temperature approach is phenomenological. We simply
insert the FS gap observed in ARPES into the band dispersion, and calculate
the responses of the emerging subbands. The experimental pseudogap is thus
mimicked by a true, but $k$-dependent, gap. Furthermore, the quasiparticles
with gapped dispersion are non-interacting. Both are standard features of
zeroth-order low-temperature calculations. We are interested in the bare
responses of the reconstructed FS, obtained at a given chemical potential.

Fig.~\ref{recFS} shows the principal responses with a small constant gap, used
for pedagogical purposes. The chemical potential cuts across both subbands,
producing the well-known disconnected reconstructed FS. The intraband nodal
response, which is the response of the arc metal, is collinearly
incommensurate, with needle-like peak shapes, consistent with experiment.

\begin{figure}
\includegraphics[height=5cm]{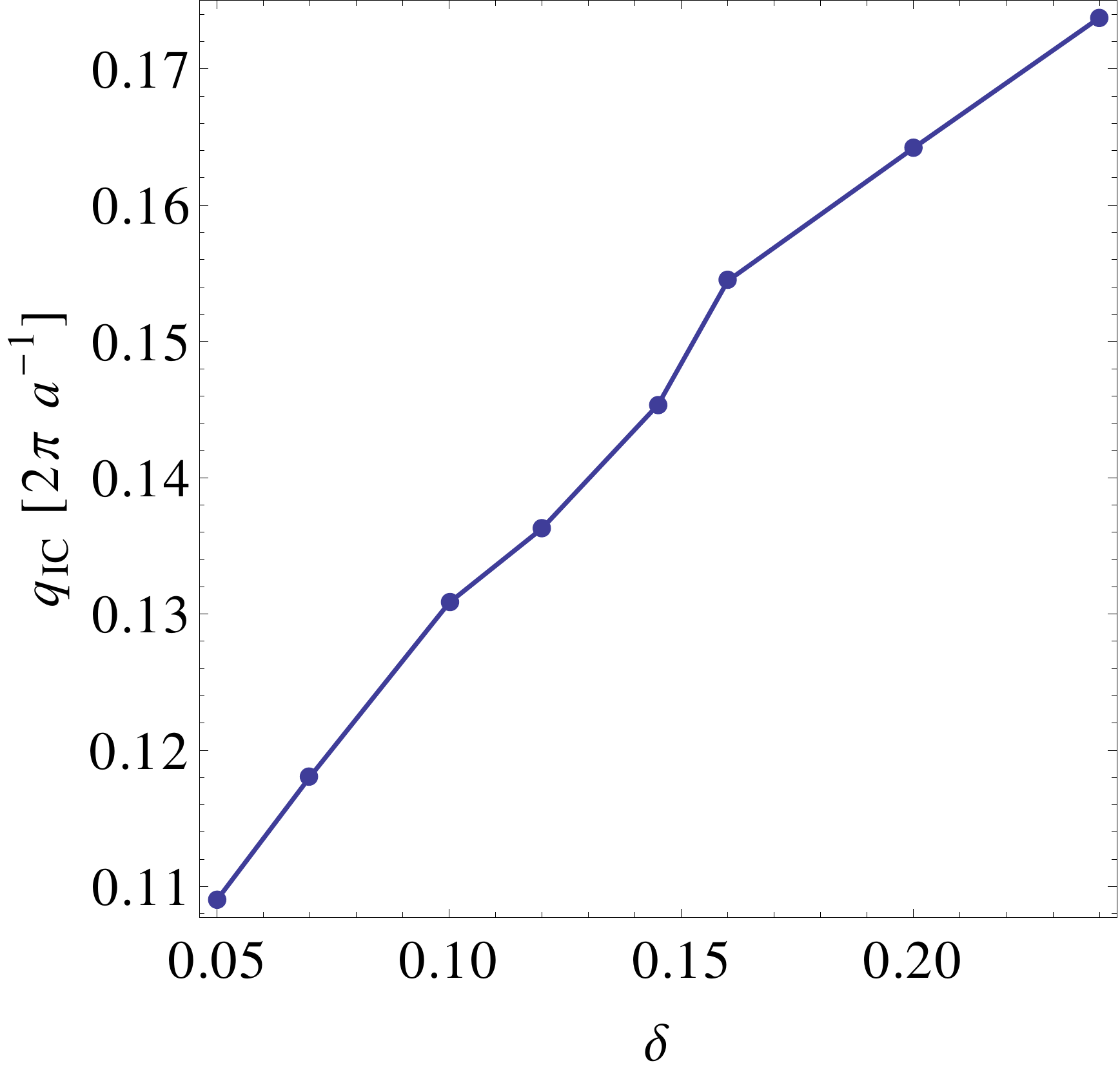}
\caption{Dependence of the collinear incommensurability away from AF on the
doping, for the free intraband arc response.}
\label{dop}
\end{figure}

Our main physical point is that the other two responses are gapped out in
real underdoped materials. They involve the antinodal segments of the FS,
which do not contribute when the observed U-shaped gap is inserted. Thus only
the nodal (arc) response survives. The doping dependence of its
incommensurability is shown in Fig.~\ref{dop}. In a wide range of doping, it
follows the experimental rule of thumb $q_{IC}\propto x$, where $q_{IC}$ is
the incommensurability away from AF, while $x$ is the hole concentration away
from half-filling, i.e.\ Ba or Sr concentration in La$_2$CuO$_4$. On the other
hand, it is fairly independent of gap size $\Delta$ as soon as the gap exceeds
the temperature, which means it becomes temperature independent as soon as
$\Delta>kT$. Both the strong doping and weak temperature dependence are
consistent with experiment.

The collinear incommensurate responses have thus been identified as the free
response of the low-temperature nodal metal. The band composition has been
crucial to arrive at this interpretation, because it underlies the microscopic
scattering which pushes the high-temperature (strongly interacting) response
to commensurability, as the copper-oxygen-hybridized holes scatter on the
copper sites. The incommensurate response of the low-temperature arc
quasiparticles is experimentally not strong enough to trigger a transition in
turn, otherwise the arc would not be conducting in the normal state. This is
consistent with our treating them as non-interacting.

\subsection{The central peak}

The calculated high-temperature commensurate response which triggers the
transition is not dispersive for energies $\omega\ll kT$. This indicates that
the transition does not proceed by a Kohn-like anomaly in the paramagnon
spectrum, but by a central peak. Such a central peak has just been reported by
J. Tranquada at this conference~\cite{Xu13}, as ``gapless magnetic
excitations.'' It appears~\cite{Xu13} that such a peak has been noticed in
earlier work, where it was interpreted as a gap with unexpected behavior,
becoming smaller with underdoping~\cite{Chang07}. We note that Fig.~2 of
Ref.~\cite{Chang07} shows that the magnetic field dependence of elastic and
inelastic neutron scattering is the same for small energy transfers
$\omega<1.5$~meV, meaning that the width of the central peak is about 15~K.

\section{Arc conductivity}

\begin{figure}
     \includegraphics[width=9pc]{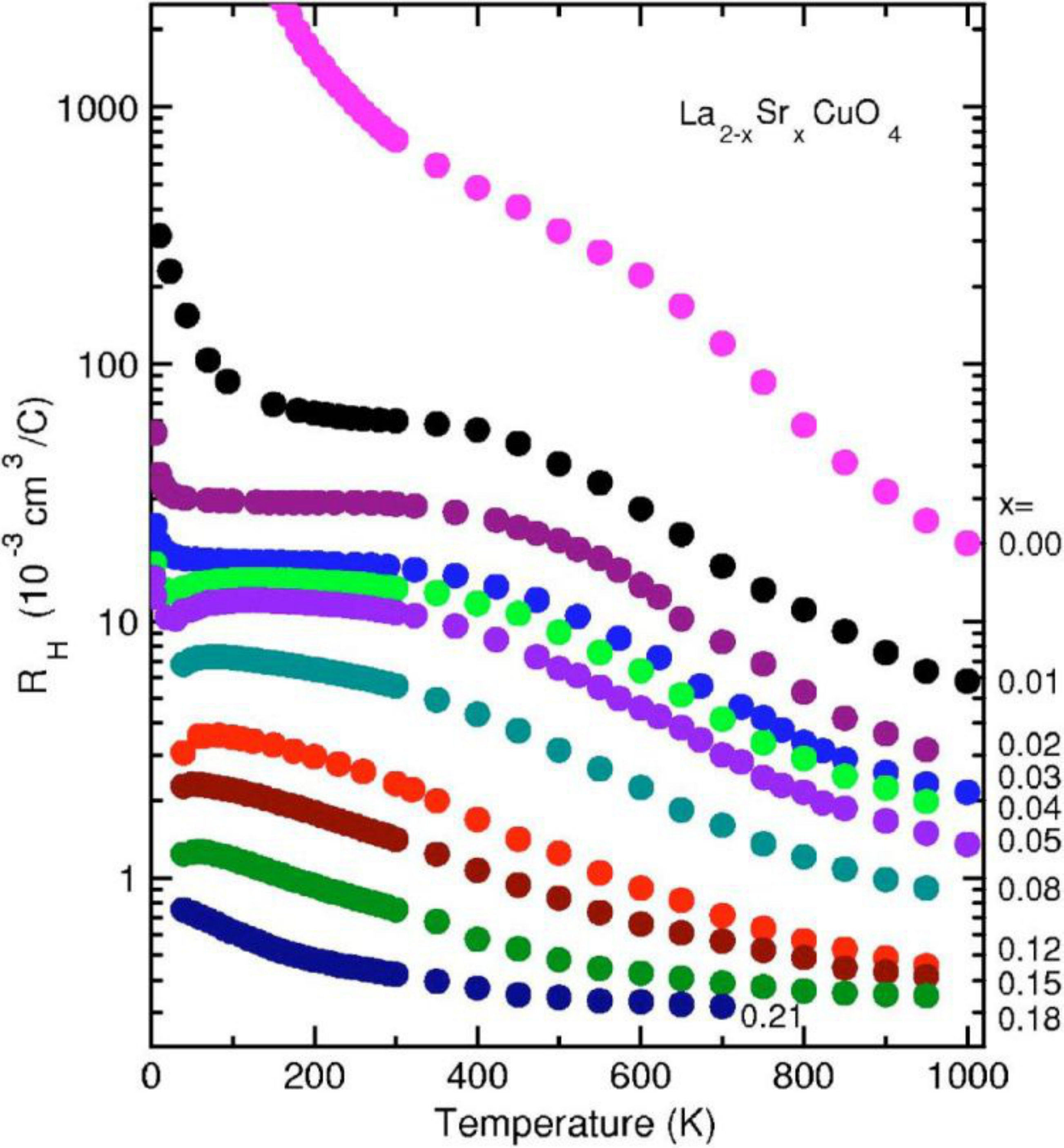}
     \includegraphics[width=9pc]{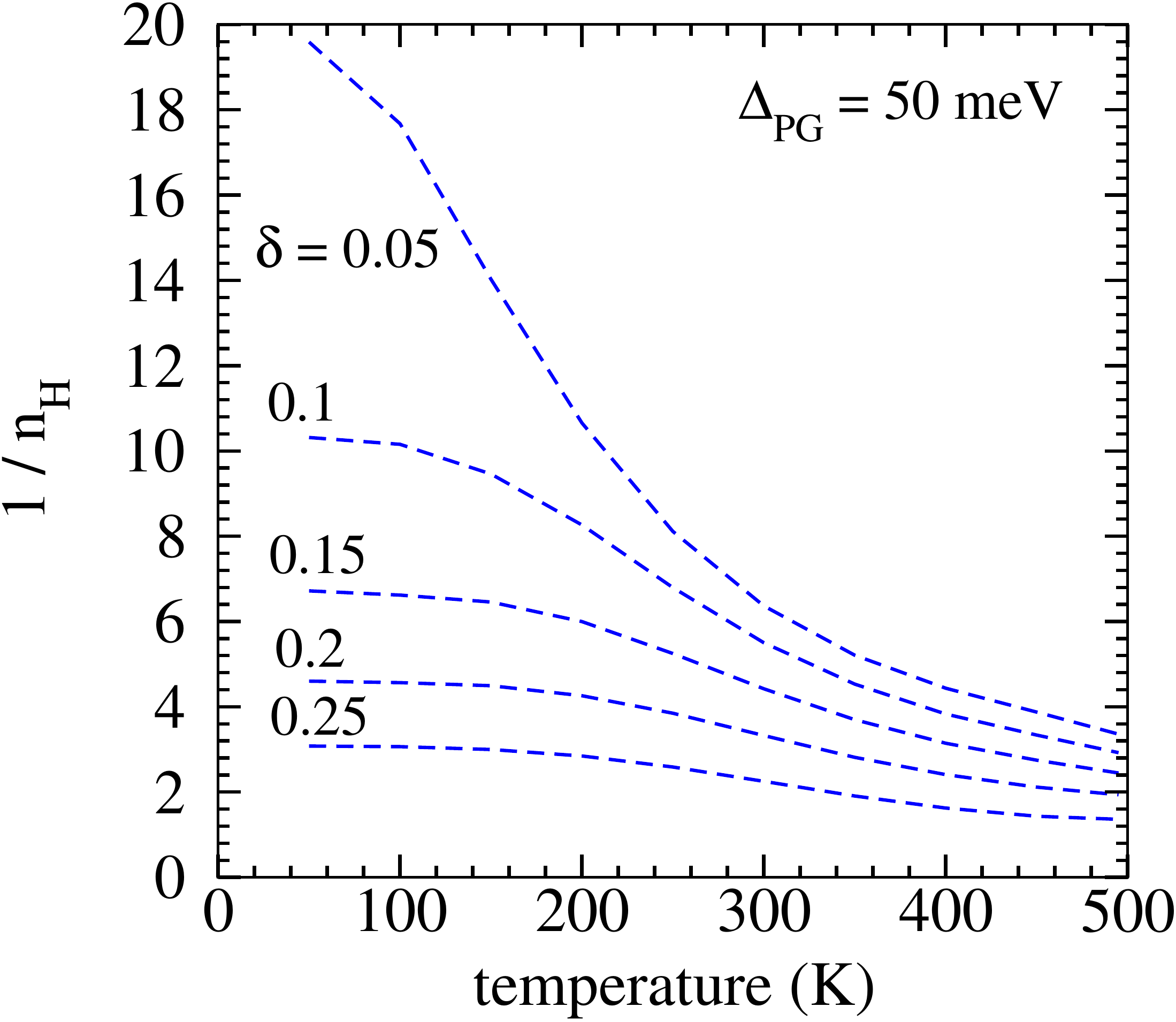}
     \includegraphics[width=9pc]{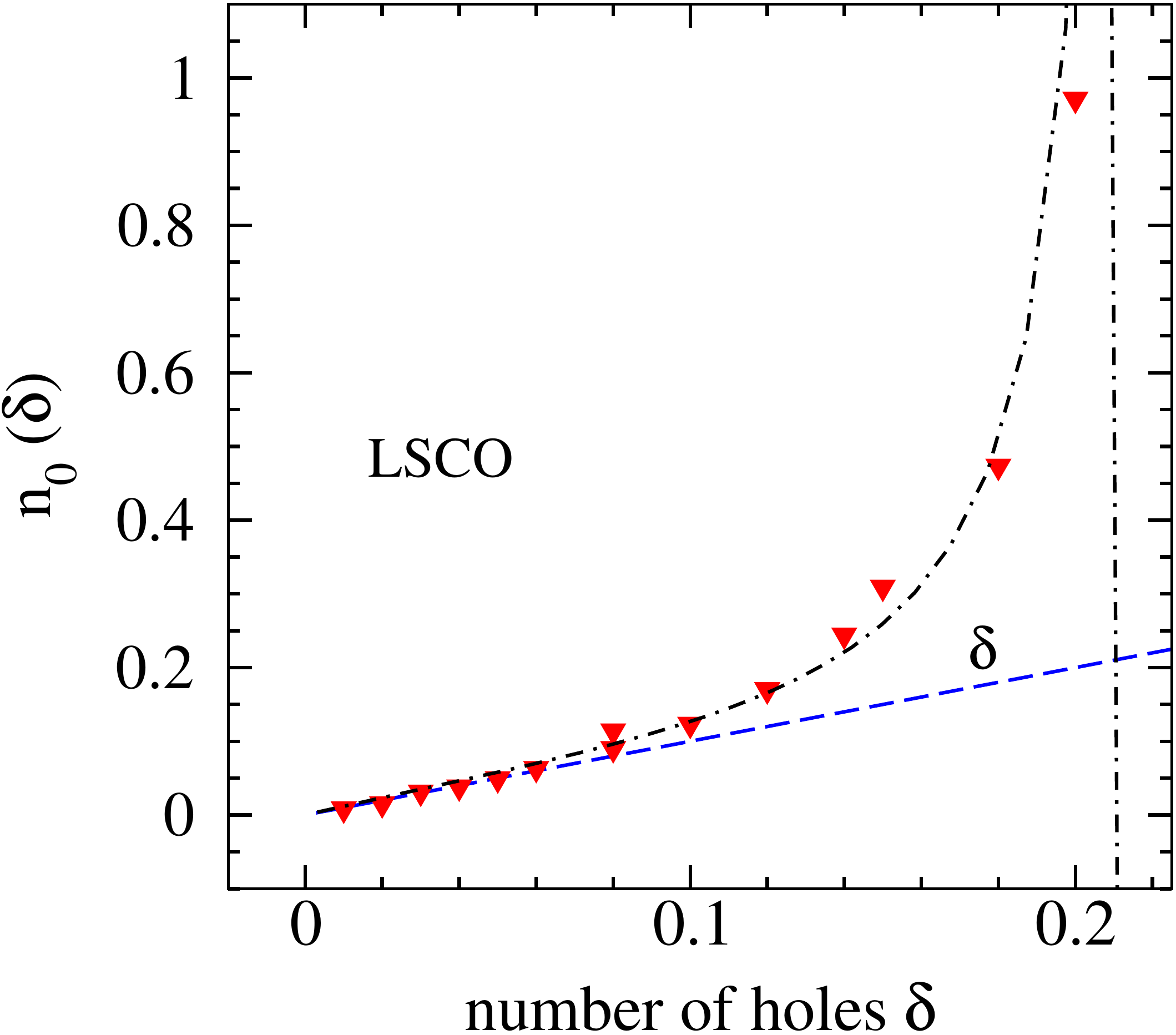}
\caption{Left: Measured Hall coefficient $R_H=1/(ecn_H)$ in LSCO~\cite{Ono07}.
Center: Eq.~(\ref{eqnH}).
Right: the FS contribution $n_0(\delta)$.}
\label{nH}
\end{figure}

The low-temperature quasiparticles invoked in the explanation of the magnetic
responses are not really non-interacting. Their residual interactions create a
pseudogap instead of the true gap used above, which means that the
quasiparticle scattering contribution to the resistivity is large, and has
indeed the expected Fermi-liquid $T^2$ dependence in the single-plane mercury
and underdoped YBCO compounds. The pseudogap scale $\Delta_{PG}$ is visible in
the temperature dependence of the Hall number $n_{\rm H}=1/(ecR_{\rm H})$ as
an activation term~\cite{Gorkov06,Kupcic07,Ono07},
\begin{equation}
n_{\rm H} =\frac{n^{\rm eff}_{xx}n^{\rm eff}_{yy}}{n^{\rm eff}_{xy}} 
\approx n_0(\delta) + n_1 {\rm e}^{- \beta
\Delta_{\rm PG}(\delta)} + \ldots
\label{eqnH}
\end{equation}
which reproduces the rapid falling-off of the Hall coefficient, as shown in
Fig.~\ref{nH}. To reproduce the doping dependence, one also needs to take into
account that the effective Hall number diverges when $n^{\rm eff}_{xy}$, which
is like a net curvature of the Fermi surface, crosses zero. This crossing
point is sufficiently close to the physical dopings that the term
$n_0(\delta)$ is strongly non-linear, as also shown in the figure.

Neither of the above effects can be understood from an effective-mass point of
view. The effective mass tensor
\begin{equation}
\left(\frac{m}{m^*}\right)^L_{\alpha\beta}=
\frac{m}{\hbar^2}\frac{\partial^2\varepsilon_L}{\partial k_\alpha\partial
k_\beta}
\end{equation}
does not naturally appear in the full quantum expression for the diagonal
effective number,
\begin{equation} n^{\rm
eff}_{\alpha \alpha} = -\frac{m}{V} \sum_{L{\bf k} \sigma} 
\left[v^L_\alpha(\bf k)\right]^2 \frac{\partial f_L({\bf k})}{
\partial\varepsilon_L({\bf k})},
\end{equation}
where $v^L_\alpha(\bf k)$ is the carrier group velocity, and $\partial
f/\partial\varepsilon_L$ is peaked at the FS, correctly suppressing
contributions from carriers deep below it. True, this expression can be
formally manipulated to give a Drude-like form, which appears to sum
contributions from all carriers,
\begin{equation} n_{\alpha \alpha}^{\rm eff} 
= \frac{1}{V} \sum_{L{\bf k} \sigma}
\left(\frac{m}{m^*}\right)^L_{\alpha\alpha} f_L({\bf k}),
\end{equation}
but this only makes sense at the bottom of the band, where the non-diagonal
term $n^{\rm eff}_{xy}$ in the denominator, which has no effective-mass
interpretation, is quiescent. In the cuprates, which are near half-filling,
the ``non-Fermi liquid'' nature of the carriers observed in Fig.~\ref{nH} is
primarily due to the failure of this effective-mass semiclassical limit, not
of the Fermi liquid concept itself. On the other hand, the Hall number
calculated in this simple band picture needs to be divided by about $4$ in
order to reproduce the absolute values of the conductivities. This is a
strong-coupling effect, by which the mobile carriers spend a large amount of
time in localized states, because of the waiting induced by the infinite $U$,
as explained above. However, the factor of four is not a direct measure of the
size of the strong-coupling effect. A careful analysis shows that the
strong-coupling renormalization factors generally appear squared, so that a
rough estimate of the localized component is rather one-half, further subject
to parameter regimes in the three-band model, which distribute the charge
between copper and oxygen.

\section{Ionicity vs. covalency}

The success of the quantum Fermi liquid is qualified by the important
limitation, that the doped holes spend only part of their time conducting, and
the rest in localized states. Technically, the quasiparticle weight is not
fully unity. Physically, bonding in every real material is on a continuous
scale between the ionic and covalent limits, and the idea that some orbitals
are exclusively localized, while others are exclusively conducting, is an
idealization, valid perhaps for the first-column metals, but certainly not for
doped transition-metal oxides. Effects of the ionic background in the metallic
behavior are the true ``strong-coupling'' effects in the cuprates. These are,
however, least visibile at the FS itself, if the latter is simply defined as
the point where some intensity is cut off by the chemical potential, because
the crossing point in the zone may differ very little from the one in the
corresponding non-interacting FS, even if the intensity profile (EDC) is quite
different than that of a quasiparticle. Such an ``experimentally defined'' FS
therefore provides a useful insight into the large background scales which
define the effective tight-binding parameters.

\subsection{Ionic effects on the site energies}

An early question in high-Tc cuprates was how the dopand charge ended in the
plane~\cite{Mazumdar89}. One possible mechanism is covalent, by which it is
discharged into the plane as if by a conducting wire. The other, advocated in
Ref.~\cite{Mazumdar89}, is ionic, by which it affects the chemical balance in
the plane by Coulomb fields, so that the plane moves from the ionic to the
covalent limit, effectively self-doping. In the tight-binding model, this is
reflected as a change in the site-energy splittings, which gives a simple
explanation, why the evolution of the FS with doping cannot be fitted with
rigid bands.

Typically, hole dopands are introduced into the cuprates in one of two ways.
One is to replace the out-of-plane metal anions, as in LSCO, where Sr replaces
La. The alternative is to introduce oxygens into interstital positions, which
are presumably cationic, like the stoichiometric oxygens. As an elementary
test of the doping mechanism, we investigate Hartree-Fock (HF) band fits to
the FS series in LSCO and single-plane Bi2201. We use a four-band model, in
which a copper 4s orbital is added to the usual three bands, because that is
the minimal model with chemically realistic parameter ranges; further
down-folding to the three-band model requires artificially large values of the
oxygen-oxygen hopping $t_{pp}$~\cite{Pavarini01}.

\begin{figure}
     \includegraphics[width=15pc]{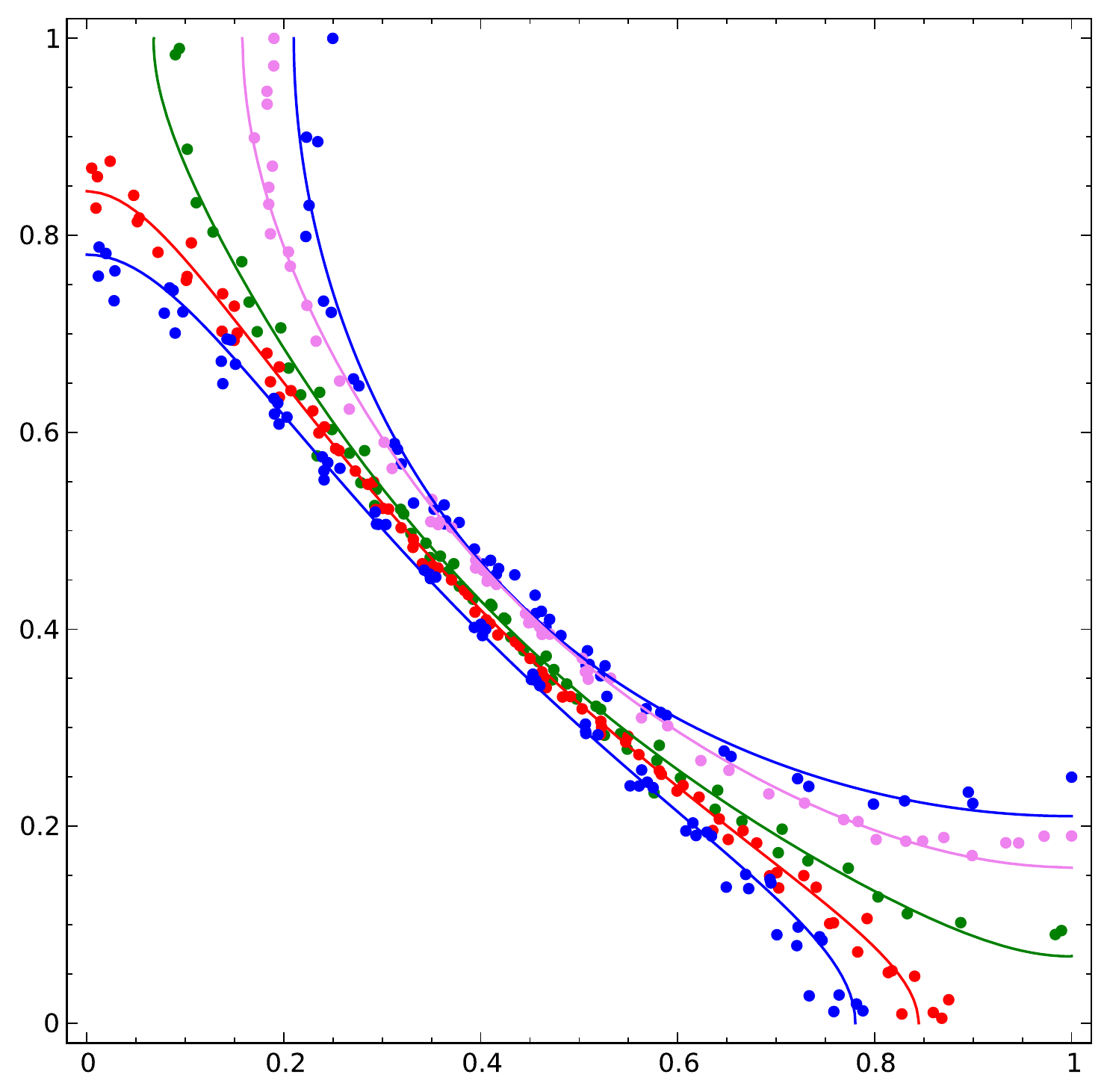}
     \includegraphics[width=15pc]{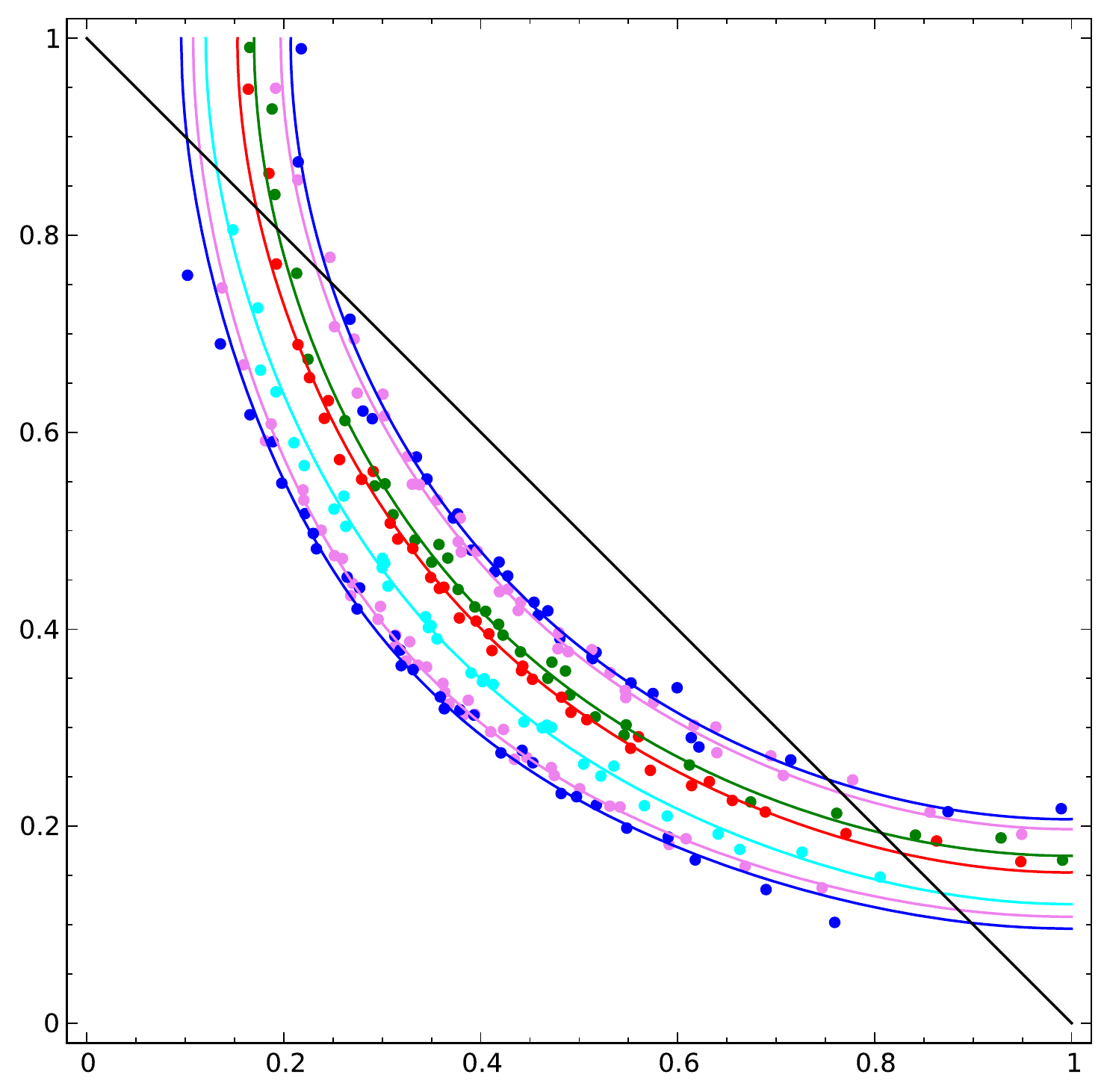}
     \includegraphics[width=12pc]{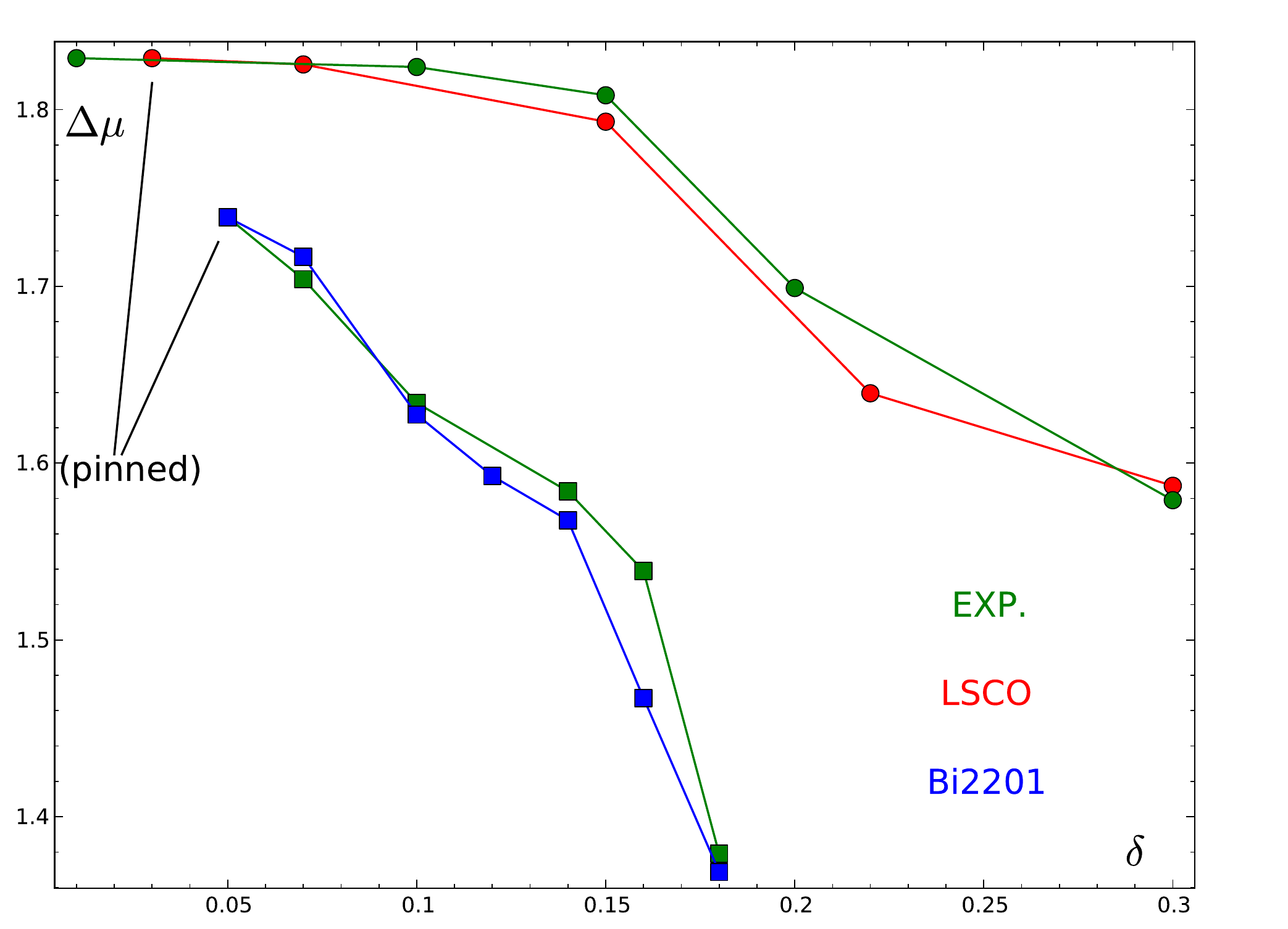}
\caption{FS fits with a four-band model for LSCO (left), Bi2201 (center), and
the corresponding chemical potentials (right). The calculated chemical
potentials have been shifted (``pinned'') to their measured values at lowest
doping.}
\label{fsfit}
\end{figure}

Fig.~\ref{fsfit} shows the Fermi-surface fits achieved in the model. Parameter
evolution has been constrained principally to the energy of the copper
4s-orbital. A further constraint was to fit the measured positions of the
absolute chemical potential within each species. This could also be achieved
at all dopings by varying only the 4s level energy, except around $15$\%
doping, where a small increase in the copper-oxygen overlap $t_{pd}$ was
required.

\begin{figure}
\includegraphics[width=12pc]{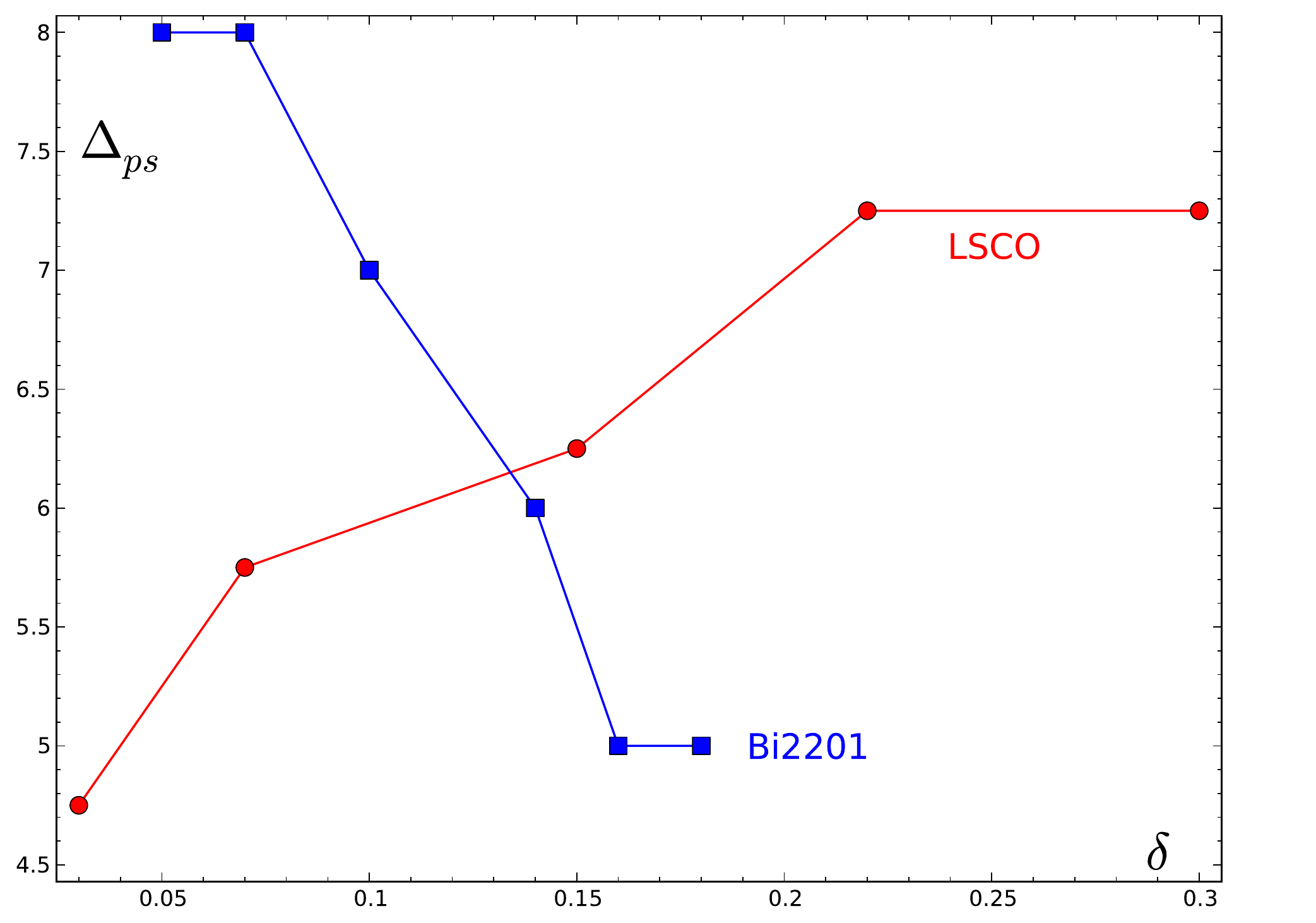}
\includegraphics[width=12pc]{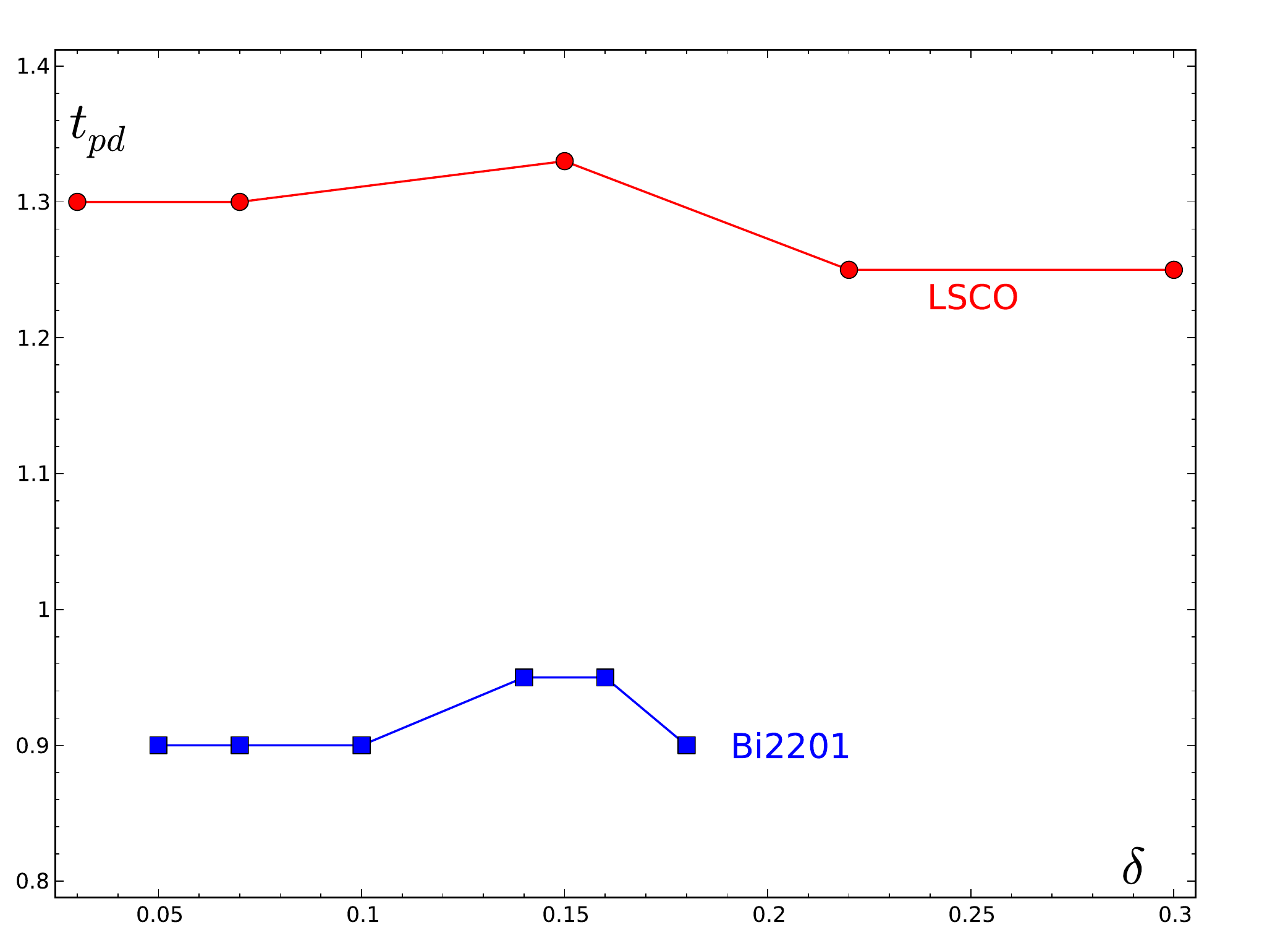}
\caption{Parameter evolution in the four-band model corresponding to the fits
in Fig.~\ref{fsfit}. All other parameters are kept fixed.}
\label{paramevol}
\end{figure}

The parameter evolution is summarized in Fig.~\ref{paramevol}. The most robust
finding is that the change of the 4s-orbital energy is in the opposite
direction for the two classes of materials. This is simply interpreted as
reflecting the different ionic positions of the dopands, supporting in turn
the ionic doping mechanism. The large overall variation of $\varepsilon_s$ is
due to the fact that it is the only parameter varied, while the comparatively
small non-uniformity in $t_{pd}$ cannot be interpreted at this level of
analysis, because other similarly small scales, such as the kink scale, are
not included. Notably, the 4s-orbital energy is the only site energy whose
variation alone can account for the whole FS evolution and observed chemical
potential shift in both materials. The \emph{increase} in the splittings
$\Delta_{ps}$ and $\Delta_{ds}$ with doping in LSCO accounts for the peculiar
hockey-stick shaped shift in the chemical potential, and its overally smaller
range than in Bi2201, because the rigid-band-filling trend of the chemical
potential runs counter to the trend determined by the increase of the level
splittings. In the interstitially doped Bi2201, both trends are in the same
direction.

\subsection{Ionic effects on extended states}

Violations of the Luttinger sum rule have been observed~\cite{Hashimoto08} in
the experimental FS crossings of the previous section. They are small relative
to the experimentally defined FS, and for the reasons outlined above, such a
small shift is presumably influenced by any number of interacting scales,
large and small. Nevertheless, an important qualitative trend is observed in
the data, namely that the violations change sign, from under- to
over-compensated, at around $10$--$15$\% doping.

\begin{figure}
\includegraphics[width=12pc,angle=-90]{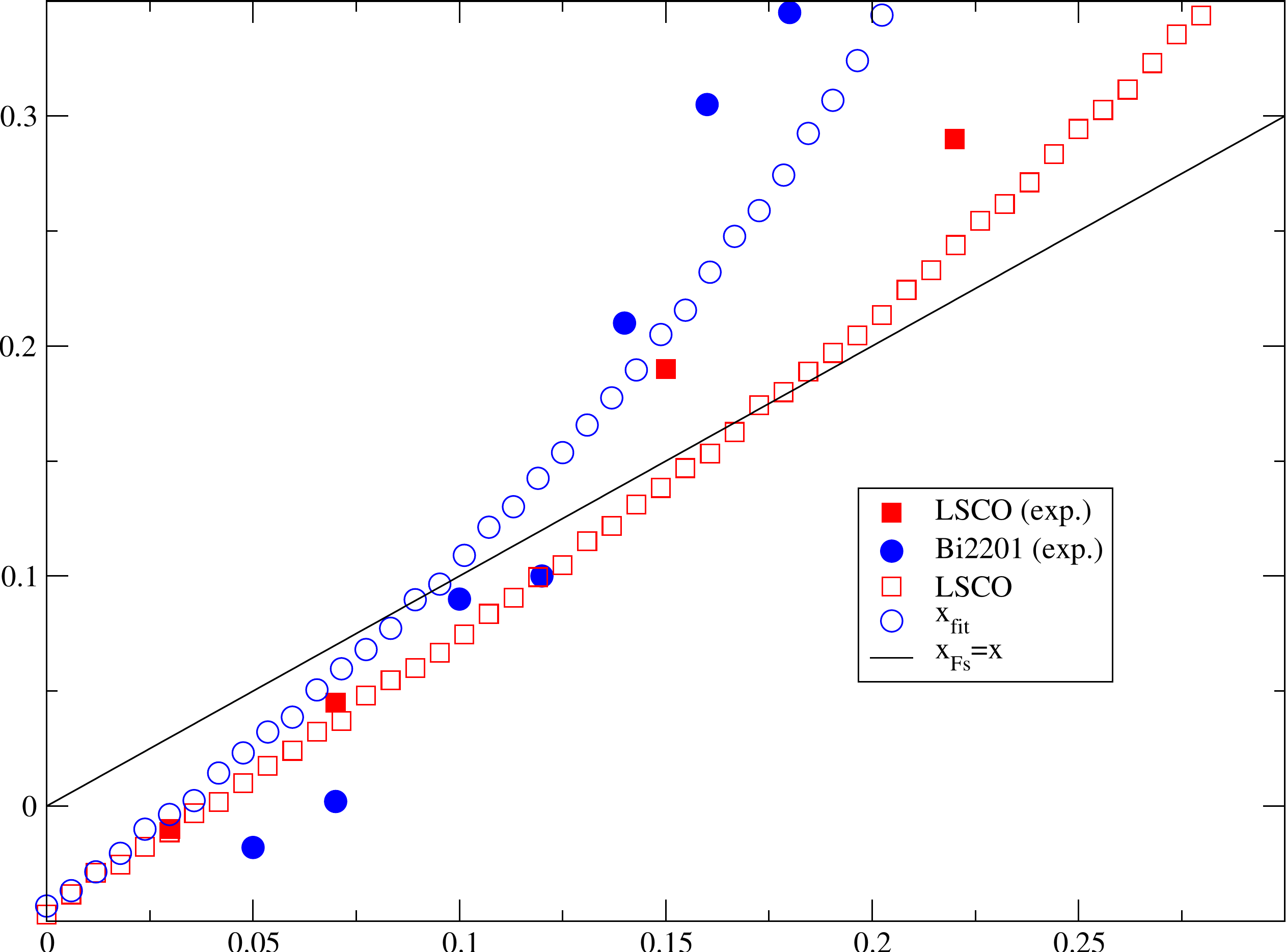}
\caption{Violations of the Luttinger sum rule.}
\label{luttsum}
\end{figure}

The same strong-coupling theoretical framework as for the magnetic responses
can describe violations from the Luttinger rule. The scattering of the band
states induces a transfer of spectral weight into incoherent localized states
on the copper sites, away from the Fermi energy. With doping, there is both a
redistribution between localized and itinerant states, and between the copper
and oxygen orbitals. The latter also affects the overall weight of the states,
because the blocking by the infinite $U$ removes some states to infinity as
the copper level is filled. In a wide range of parameters, the net effect is a
change in sign of the Luttinger-rule violation roughly, and robustly, where it
is observed experimentally, as shown in Fig.~\ref{luttsum}. We conclude that
the dominant mechanism of Luttinger-rule violation is by the large-$U$ removal
of spectral weight away from the Fermi level, with quantitative details left
to be explained by the lower-energy mechanisms, such as the nodal kink, not
included in the calculations depicted in the figure.

\section{Conclusion}

The cuprates lie in a crossover region between the ionic and covalent limits.
The ionic aspect is reflected by the doping affecting the underlying
tight-binding parameters of the metal directly, in addition to changing the
concentration. Close to the Fermi surface, however, there is every indication
that the metal behaves as a Fermi liquid out of a reconstructed Fermi
surface~\cite{Mirzaei12,DoironLeyraud13}, except that semiclassical intuitions
based on the effective-mass picture are misleading. We have presented three
distinct lines of evidence in support of that view.

We can account for the observed incommensurate collinear magnetic responses as
the free responses of the arc metal, after it has gone through a magnetic
transition via a central peak mechanism. The doping and temperature trends in
the  Hall conductivity of the arc metal closely conform to experiment with the
single extraneous assumption, that the overall spectral weight is
experimentally reduced due to localized states. The non-rigid-band FS
evolution with doping is understood through Coulomb effects of the dopands on
the orbital splittings in the plane. Finally, the comparatively small
Luttinger-sum violations are a direct consequence of the on-site repulsion
visible at the Fermi level itself, but as an overall (bulk) effect do not
invalidate the quasiparticle picture of the carriers, although their spectral
weight is significantly reduced.

Our point of view has received additional experimental support at this
conference. The identification of the central peak in the magnetic
response~\cite{Xu13} has been discussed above. The extension of the SC region
to zero doping in PLCCO upon transformation from the T to the T'
lattice~\cite{Adachi13}, similar to the one in NCCO thin
films~\cite{Tsukada05}, is consistent with our understanding of the effects of
out-of-plane dopand ions. We concur~\cite{Adachi13} that the removal of the
apical oxygens has caused a strong reduction of the in-plane copper-oxygen
splitting. Subsequent charge redistribution between coppers and oxygens
eliminates superexchange, which requires the oxygens to be empty, so that
static AF disappears.

An essential ingredient of the picture we draw of all the cuprates is to
distinguish copper and oxygen sites. Their relative roles in the SC response
remain to be elucidated. A prerequisite for such a step is the proper
disentanglement of low energy magnetic and high energy Coulomb effects, which
was the principal theme of the present work, and could only be accomplished in
a multi-band setting.

\acknowledgments

Conversations with J.~Tranquada and source files of published data provided by
A.~Fujimori are gratefully acknowledged. This work was supported by the
Croatian Government under Project No.~119--1191458--0512.

\end{document}